\documentclass[english,aps,pre, preprint, onecolumn]{revtex4}
\usepackage[T1]{fontenc}
\usepackage[latin9]{inputenc}
\usepackage{float}
\usepackage{amsmath}
\usepackage{graphicx}
\usepackage[caption=false]{subfig}
\makeatletter
\@ifundefined{textcolor}{}
{%
 \definecolor{BLACK}{gray}{0}
 \definecolor{WHITE}{gray}{1}
 \definecolor{RED}{rgb}{1,0,0}
 \definecolor{GREEN}{rgb}{0,1,0}
 \definecolor{BLUE}{rgb}{0,0,1}
 \definecolor{CYAN}{cmyk}{1,0,0,0}
 \definecolor{MAGENTA}{cmyk}{0,1,0,0}
 \definecolor{YELLOW}{cmyk}{0,0,1,0}
}

\makeatother

\usepackage{babel}
\begin{document}

\preprint{\%This line only printed with preprint option}

\title{Direct observation of the effects of spin dependent momentum of light in  optical tweezers}

\author{Debapriya Pal$^1$}
\author{Subhasish Dutta Gupta$^{2,3}$}
\author{Nirmalya Ghosh$^{1\ast}$}
\author{Ayan Banerjee$^{1\ast}$}
\email{**ayan@iiserkol.ac.in}
\email{*nghosh@iiserkol.ac.in}
\affiliation{$^1$ Department of Physical Sciences, Indian Institute of Science Education and Research Kolkata, Mohanpur 741246, India}
\affiliation{$^2$ School of Physics, Hyderabad Central University, Hyderabad 500046, India}
\affiliation{$^3$ TIFR Centre for Interdisciplinary Sciences, Hyderabad 500107, India}

\begin{abstract}
We demonstrate that tight focusing of a circularly polarized Gaussian beam in optical tweezers leads to spin-momentum locking -  with the transverse spin angular momentum density being  independent of helicity, while the transverse  momentum $($Poynting vector$)$ becomes helicity dependent. Our theoretical calculations, numerical simulations, and experiments  reveal that the presence of a stratified medium in the path of the trapping beam significantly enhances the magnitude of transverse momentum in the radial direction with respect to the beam axis, and likewise, also leads to high off-axial intensity. This overlap  allows us to experimentally observe the circular motion of a birefringent particle, trapped off-axis, in response to an input circularly polarized fundamental Gaussian beam carrying no intrinsic orbital angular momentum. The circular motion is dependent on the helicity of the input beam, so that we can identify it to be the signature of the elusive Belinfante spin in propagating light beams obtained in our optical tweezers setup. Our work can be  extended to higher-order beams carrying intrinsic orbital angular momentum leading to simple routes of achieving complex particle manipulation using optical tweezers.
\end{abstract}
\maketitle

\section{Introduction}

Light carries both orbital and spin angular momentum. The Poynting vector -  considered to be the vector representative of the flow of energy - has contribution from both the canonical and spin part of the momentum. The spin contribution in the Poynting vector ($\mathbf{P}$, or total momentum), introduced by Belinfante through the equation $\mathbf{P}=\mathbf{P}_{o}+\mathbf{P}_{s}$ \cite{belinfante1940current, bliokh2014extraordinary} - where $\mathbf{P}_{s}$ represents the spin momentum - is rather enigmatic, since the term $\mathbf{P}_{s}$ - while being responsible for spin angular momentum - does not contribute to the energy flow and is therefore considered to be a virtual quantity. On the other hand, $\mathbf{P}_{o}$ represents canonical momentum which is  responsible for generating orbital angular momentum (OAM) $\mathbf{l}$ ($\mathbf{l}=\mathbf{r} \times \mathbf{P}_{o}$, where $\mathbf{r}$ is the distance from the beam axis), that is directly manifested in experiments by the rotation of mesoscopic particles about the beam axis in optical tweezers \cite{halina1998, Nieminen2008}. The question thus arises whether the manifestation of the elusive Belinfante spin momentum can be experimentally extracted by similar means. 

It has recently been observed that a  longitudinal component of the field - phase-shifted with respect to the transverse component - plays a major role in the appearance of spin (polarization) dependent transverse momentum and spin (polarization) independent transverse spin angular momentum (TSAM)  \cite{bliokh2014extraordinary, aiello2015transverse,bliokh2015quantum,bekshaev2015transverse, gupta2015wave,eismann2019spin}. This particular feature is well known as spin momentum locking in condensed matter physics in the context of topological insulators \cite{hasan2010colloquium}, where  special states exist at the outermost surface of the insulator which fall within the bulk energy gap and permit surface metallic conduction. The carriers in these surface states are observed with their spin locked at a right-angle to their momentum (spin-momentum locking) \cite{hsieh2008topological}. In optics, this feature is manifested as the transverse component of the Poynting vector - which represents the flow of momentum - being dependent on helicity (spin) of the beam. In case of evanescent fields, such non-trivial structures of spin and momentum density have already been reported \cite{bliokh2014extraordinary,van2016universal}. In fact, such a transversely spinning electric field arising in the case of transverse SAM of light, and resembling the spinning movement of the spokes of a rolling bicycle wheel, has recently been experimentally achieved \cite{bauer2016optical}. It has also been shown that the general solution of Mie scattering from a spherical particle, which has a phase-shifted longitudinal component indeed has the helicity dependent transverse component of Poynting vector (generally addressed as `transverse (spin) momentum') and helicity independent transverse spin angular momentum density \cite{saha2016transverse, saha2018effects, gupta2015wave}. Thus, keeping in mind that a tightly focused Gaussian beam has a longitudinal field component which is phase shifted from the transverse components, the question that naturally arises is whether  such a beam also contains these interesting and exotic properties. Now, it is interesting to note that while a few practical applications such as nano-displacement probes \cite{rodriguez2010optical}, or generation of optical vortices \cite{brasselet2009optical} have been developed, in most cases the effects of spin-orbit interaction (SOI) have been quite small, with the magnitude of trajectory shifts reported due to the Spin Hall Effect (SHE) of light usually being in sub-wavelength regime \cite{bliokh2015spin,KB1,rubinsztein2016roadmap}. However, in Refs. \cite{rodriguez2010optical, bliokh2011spin}, the SHE due to scattering of nano-particles was enlarged by using a imaging system with small focal length or high numerical aperture (N.A.) microscope objectives so that the extent of the SHE covered the microscope exit pupil. Recently, we showed that optical tweezers - due to the tight focusing involved - causes SOI inherently \cite{basudevpra2013}, and the use of a stratified medium in the beam path in the tweezers enhances polarization-dependent intensity distributions and SHE effects which can lead to controlled particle manipulation \cite{basudevpra2013,basudevnjp2014}. Thus, it makes sense to explore the use of optical tweezers in observing spin-momentum locking in Gaussian beams, and investigate the manifestations of Belinfante spin.

In this paper, we demonstrate that a tightly focused spin-polarized Gaussian beam indeed possesses the very same properties of spin-momentum locking that have been observed for evanescent fields and Mie scattering. We simulate the electric and magnetic field distribution for an optical tweezers configuration, and the results of our simulations clearly demonstrate the existence of spin momentum locking in such beams. Our simulations also reveal that the presence of a stratified medium in the path of the trapping beam can significantly enhance the magnitude of transverse spin angular momentum and transverse Poynting vector, which point to the possibility of observing effects of Belinfante spin experimentally. We verify this experimentally by observing spin-dependent rotational motion of a birefringent particle around the beam axis for input circularly polarized light propagating through a stratified medium.
\section{Theory}
In case of tight focusing of light, the paraxial approximation fails with the generation of a large longitudinal component of the electric field which is phase shifted from the transverse components. Therefore, we calculate the electric field distribution of the tightly focused beam passing through different refractive index layers (considering both forward and backward propagation) employing the angular spectrum method \cite{rohrbach2005stiffness} using Debye-Wolf diffraction integrals \cite{richards1959electromagnetic}. The expression of the tightly focused beam can be written in the form of a matrix equation as:
\begin{equation}
\left({\begin{array}{c}
E_{x} \\ 
E_{y} \\ 
E_{z}
\end{array} } \right)= C
\left({\begin{array}{ccc}
I_{0} + I_{2} cos 2 \psi  & I_{2} sin {2 \psi}  & 2iI_{1} cos{\psi}   \\ 
I_{2} sin {2 \psi}  & I_{0} - I_{2} cos {2 \psi}  & 2iI_{1} sin{\psi}   \\ 
-2iI_{1} cos{\psi}   & -2iI_{1} sin{\psi}   & I_{0}+I_{2}
\end{array} } \right) \vec{X}
\label{matrix}
\end{equation}
where $\vec{X}$ is the Jones vector of the input field, $I_0$, $I_1$ and $I_2$ are the Debye integrals (see Supplementary Information), and C is a constant. The beam propagation direction is in the $\mathbf{z}$, whereas $\mathbf{x}$, and $\mathbf{y}$ denote the transverse plane. The diffraction integrals account for the forward and backward propagating beams, and contain the Fresnel coefficients ($T_s,\,T_p,\,R_s $ and $R_p$) which are required to consider the effect of the stratified medium.  We keep track of the evolution of the electric field in each  refractive index layer of the stratified medium using:
\begin{equation}
\label{ts}
T_i(j,j+1)=\frac{E_{i+}^{j+1}}{E_{i+}^{j}} 
\end{equation}
\begin{equation}
\label{rs}
R_i(j,j+1)=\frac{E_{i-}^{j+1}}{E_{i-}^{j}}. 
\end{equation}  
Here, i denotes the polarization states (i.e. s-polarized or p-polarized denoted by s and p respectively), and the positive and negative signs denote the forward and backward propagating waves, respectively, and $j$ denotes a particular refractive index layer.
The output electric fields can be obtained using the Jones vector for left and right circularly polarized input light as $\vec{X}=[1~~\pm i~~0]^T$. Note that the tightly focused electric and magnetic field enjoy the following symmetry relations (in Gaussian units)  \cite{richards1959electromagnetic}:
\begin{equation}
\begin{array}{l}{\mathbf{H}_{x}=-\mathbf{E}_{y}\left(\psi - \frac{\pi}{2}\right)} \\ {\mathbf{H}_{y}=\mathbf{E}_{x}\left(\psi - \frac{\pi}{2}\right)} \\ {\mathbf{H}_{z}=\mathbf{E}_{z}\left(\psi - \frac{\pi}{2}\right)}\end{array}
\end{equation}
As circular polarized light propagates along the \textbf{z} direction, the longitudinal spin angular momentum (LSAM) is generated due to the intrinsic spin (helicity $\sigma$) of the light and may be represented as $\mathbf{S} \propto \sigma \mathbf{z}$, which can be transferred to absorbing birefringent particles to cause rotation about their centre of mass. As noted previously, the  Poynting vector or momentum density of the wave can be broken up into the canonical (or orbital, $\mathbf{P_o}$) and spin ($\mathbf{P_s}$) components. The latter can also be transferred locally to probe particles resulting in a torque $\textbf{T} \propto \mathbf{P_s}$ about the beam axis, in contrast to the SAM($\mathbf{S}$) which generates rotation around the particle center of mass. Now, the spin angular momentum density ($\mathbf{S}$) and Poynting vector ($\mathbf{P}$) in an isotropic medium, considering real $\epsilon$ and $\mu$, in S.I. units are given as (ignoring dispersion effects):
\begin{equation}
\mathbf{S}=\frac{\epsilon \mathbf{E} \times \mathbf{E}^{*}+\mu \mathbf{H} \times \mathbf{H}^{*}}{\omega\left[\epsilon|\mathbf{E}|^{2}+\mu|\mathbf{H}|^{2}\right]}
\end{equation}
\begin{equation}
\mathbf{P}=\frac{1}{2} \operatorname{Re}\left(\mathbf{E} \times \mathbf{H}^{*}\right)
\end{equation}
Using the above relations, we calculate the transverse spin angular momentum (TSAM) density  and transverse momentum (TM, or transverse Poynting vector) density for right and left circular polarized light. The expressions for right circular polarized light are:
\begin{equation}
\begin{array}{l}{\mathbf{S}_{x}=-4 i I_{1}\left(I_{0}+I_{2}\right) \sin (\psi)} \\ {\mathbf{S}_{y}=4 i I_{1}\left(I_{0}+I_{2}\right) \cos (\psi)} \\ {\mathbf{S}_{z}=-2 i\left(I_{0}^{2}-I_{2}^{2}\right)} \\\\ {\mathbf{P}_{x}=4 I_{1}\left(I_{0}+I_{2}\right) \sin (\psi)} \\ {\mathbf{P}_{y}=-4 I_{1}\left(I_{0}+I_{2}\right) \sin (\psi)} \\ {\mathbf{P}_{z}=2\left(I_{0}^{2}-I_{2}^{2}\right)},\end{array}
\end{equation}
while, those for left circular polarized light are:
\begin{equation}
\begin{aligned} \mathbf{S}_{x} &=-4 i I_{1}\left(I_{0}+I_{2}\right) \sin (\psi) \\ \mathbf{S}_{y} &=4 i I_{1}\left(I_{0}+I_{2}\right) \cos (\psi) \\ \mathbf{S}_{z} &=2 i\left(I_{0}^{2}-I_{2}^{2}\right) \\\\ \mathbf{P}_{x} &=-4 I_{1}\left(I_{0}+I_{2}\right) \sin (\psi) \\ \mathbf{P}_{y} &=4 I_{1}\left(I_{0}+I_{2}\right) \sin (\psi) \\ \mathbf{P}_{z} &=2\left(I_{0}^{2}-I_{2}^{2}\right) \end{aligned}
\end{equation}
Clearly we can see that under the change of helicity from +1 to -1, the TSAM remains same, but the TM  flips direction, which confirms that the helicity-independence of the former and helicity-dependence of the latter are inherent properties of a tightly focused Gaussian beam.

As discussed earlier, $T_s,\,T_p,\,R_s\,\text{and}\,R_p$ have a crucial role to play in determining the diffraction integrals which in turn determine the final electric field and magnetic field distributions. Thus, a suitable choice of refractive indices in the stratified medium may provide us control over the Fresnel's coefficients which would determine the final distribution of the electric and magnetic field, and produce interesting effects.
\section{Numerical Simulations}
\begin{figure}[h]
    \centering
    \includegraphics[width=0.6\textwidth]{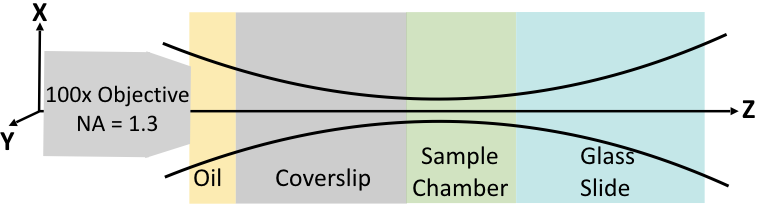}
    \caption{Schematic of our sample chamber setup. (Dimensions not to scale) The \textbf{z} - axis is the propagation direction, whereas \textbf{x} - axis and \textbf{y} - axis describes the transverse plane.}
    \label{fig:setup}
\end{figure}

We now run simulations on our experimental system (stratified medium in the path of the optical tweezers light beam) with the theoretical model developed in the previous section and observe the occurrences of diverse phenomena dealing with spin momentum locking,  transverse spin and spin momentum effects for input circular polarized Gaussian beam ($TEM_{00}$ mode) into the optical tweezers. The laser beam of wavelength $1064\,nm$ is incident on the $100X$ objective of numerical aperture 1.4, followed by: a) an oil layer of refractive index (RI) 1.516, b) a 160 microns thick cover-slip having refractive indices 1.516 and 1.814 (note that the case where the RI = 1.516 is henceforth referred to as the `matched condition' since this the most general condition employed in optical tweezers experiments so as to minimize spherical aberration effects in the focused beam spot), c) an aqueous solution chamber having refractive index of 1.33 with a depth of 35 microns, and finally d) a glass slide of refractive index 1.516 whose thickness we consider to be  semi-infinite (around 2000-3000 microns, being very large considered to other dimensions) as shown in Figure.$\,$\ref{fig:setup}.
We first define a specific coordinate system for our simulation in which the $z = 0$ is taken at $5\,\mu m$ inside the sample chamber near the cover-slip. Considering this as the origin and with the dimensions mentioned above, the boundary for oil and objective is at $-170\,\mu m$, oil and cover-slip at $-165\,\mu m$, cover-slip and sample chamber at $-5\,\mu m$ and sample chamber and glass slide at $+30\,\mu m$. 

\begin{figure}[h]
    \centering
    \includegraphics[width=0.6\textwidth]{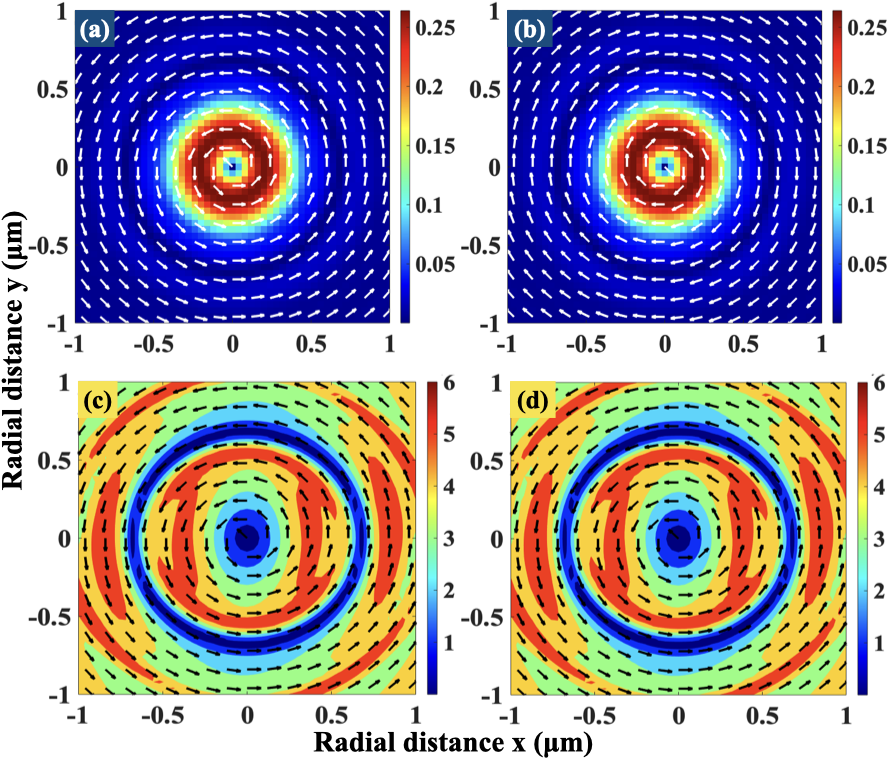}
    \caption{Transverse component of Poynting vector for a) left circularly polarized beam; b) right circularly polarized beam; Transverse component of spin angular momentum for c) left circularly polarized beam; d) right circular polarized beam. All plots are at the focus of the beam for a cover slip RI of 1.516.}
    \label{fig:SpinMomentum}
\end{figure}

We use both left circular and right circular light as input and compare the results with respect to spin-momentum locking. Figure\,\ref{fig:SpinMomentum}a) and \ref{fig:SpinMomentum}c) display the TSAM and TM for a left circularly polarized beam, while Figure\,\ref{fig:SpinMomentum}b) and \ref{fig:SpinMomentum}d) displays the same for right circularly polarized beam at the beam focus. It is clear from the quiver plots that the TM vector flips sign whereas the TSAM is independent of helicity of the input beam. This establishes our claim of the existence of spin momentum locking in tightly focused Gaussian beams.
With the change in the RI contrast within different layers, the Fresnel coefficients (which determine the nature of the Debye integrals) change, which would affect the intensity distribution, as well as the SAM density and Poynting vector. However, in order to probe effects of the angular momenta on probe particles, we need to trap them first - which makes the field intensity distribution at the sample region of optical tweezers crucial. 

We now shift our attention to this problem, and quantify the intensity profile in the radial direction inside the sample (water in our case) for cover-slips of different RI and also investigate whether the off-axis intensity is sufficient enough for trapping and rotating micron sized particles given the magnitude of the TM at the same location. We do not consider scattering effects from a trapped particle, since the scattering (which itself is very small compared to the transmitted field due to the size of the particle) is predominantly in the forward direction, and does not contribute to the transverse field anyway.\\
\begin{figure}[h]
    \centering
    \includegraphics[width=0.65\textwidth]{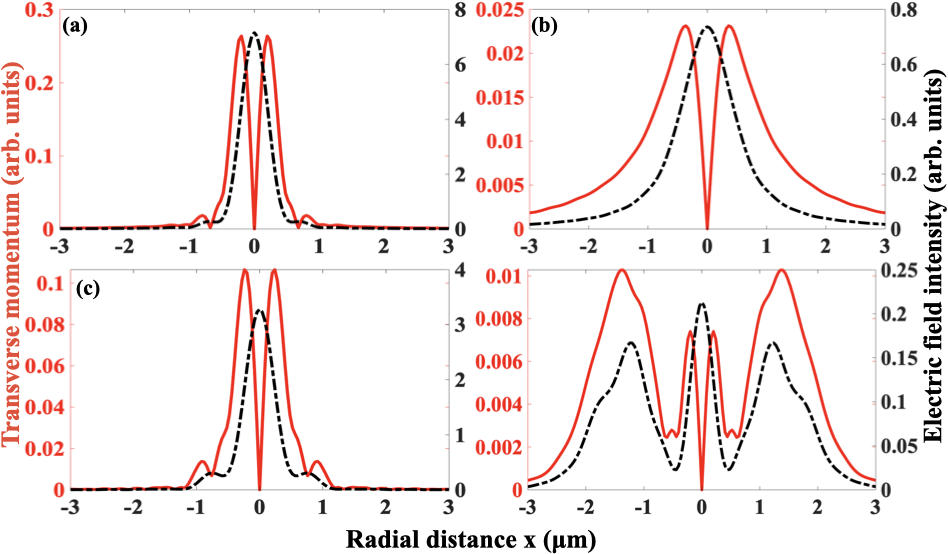}\caption{Radial intensity(black) and transverse spin momentum(red) plot for cover-slips of RI $1.516$ at a) focus; b) 2 $\mu m$ away from focus; and  RI $1.814$ at (c) focus; (d) 2 $\mu m$ away from focus}
    \label{fig:ElectricSpin}
\end{figure}
\newline
We plot Fig.\,\ref{fig:ElectricSpin}a) and \ref{fig:ElectricSpin}b) for the matched condition (cover-slips of RI $1.516$) at focus and $2 \mu m$ axially (Rayleigh range using Gaussian approximation is $\sim$ 175 $\mu m$)  away from the focus, respectively. The intensity distribution is clearly Gaussian with negligible side lobes. However, since we are working with a fast diverging Gaussian beam, the beam intensity reduces rapidly. The side lobes also disappear as the beam expands in the radial direction when we move $2 \mu m$ beyond the focus as shown in Fig.\,\ref{fig:ElectricSpin}b). It is also interesting to note that the TM -\,given by the red line\,- is zero at the beam center, and increases slightly off-axis, where the field intensity has fallen substantially. Thus, the possibility of  observation of any effect of the TM on trapped particles is rather low, which is perhaps the main reason why this has not been observed in conventional optical tweezers.\\
\newline
However, when we increase the RI of the cover-slips to $1.814$ (mismatched condition), the intensity distribution is inhomogeneous, and side lobes are formed as shown in Fig.\,\ref{fig:ElectricSpin}c) and \ref{fig:ElectricSpin}d). Indeed, the intensity at the beam center is substantially lowered for the higher RI due to  diffraction effects (spherical aberration). The interesting point is that the TM -\,depicted by red lines\,- increases in magnitude off-axis for the increased RI. The intensity is also high here (close to that at the centre), and we now obtain regions where both the intensity and transverse spin momentum are high enough in magnitude, so that particles may be trapped, and also rotate due to the presence of sufficient TM. Thus, a particle, trapped in this configuration with a stratified medium in the path of the optical tweezers light beam may be used as a probe to observe the TM.
\begin{figure}[h]
    \centering
    \includegraphics[width=0.65\textwidth]{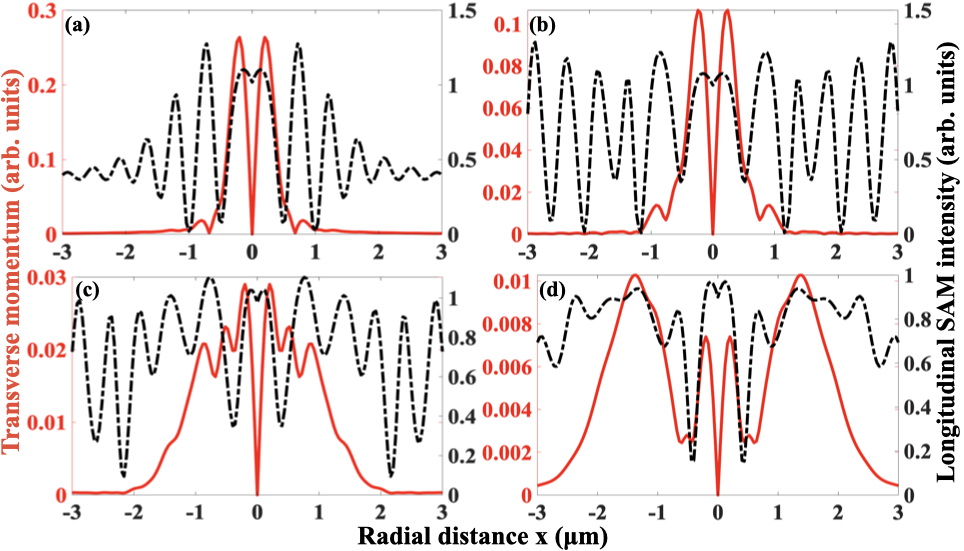}
    \caption{Longitudinal Spin Angular Momentum Intensity (black) and transverse spin momentum (red) plot for cover-slip of RI $1.516$ at (a) focus; and RI $1.814$ at (b) focus; (c) 1 $\mu m$; and (d) 2 $\mu m$ away from focus}
    \label{fig:TSMLSAM}
\end{figure}
\newline

Now, we go on to investigate the LSAM for our system, which we plot in  Fig.\,\ref{fig:TSMLSAM} for both matched (Fig.~\ref{fig:TSMLSAM}(a)) and mismatched conditions (Fig.~\ref{fig:TSMLSAM}(b) - (d)). We observe that the LSAM - depicted by black lines - is always high at the beam center, where the intensity is also maximum, which explains the routine observation of the rotation of transparent birefringent particles around their centre of mass for input circular polarization in optical tweezers \cite{halina1998}. However, at the center, the TM - depicted by red lines - is low. For different axial planes, the matched condition yields results (displayed in detail in the SI) uninteresting for experiments since the field intensity falls of rapidly in the radial direction. The mismatched case, though, needs to be considered in detail. Here, we observe that while the TM has very small radial lobes at the focus, these start becoming more significant as we move away from the focus axially. At z=1 $\mu$m, we observe the LSAM has several off-axis peaks, whereas the TM still falls off quickly from the center. However, at z=2 $\mu$m, there is a region where both TM and LSAM are high off-axis in a particular region - which along with Fig.~\ref{fig:ElectricSpin}(d) (which shows that the field intensity is also high in that region) - suggests that it may be possible to observe interesting effects in particle rotation for this situation. This is what we experimentally pursue, as we describe in the next section.
\section{Experiment}
\begin{figure}[h]
    \centering
    \includegraphics[width=0.75\textwidth]{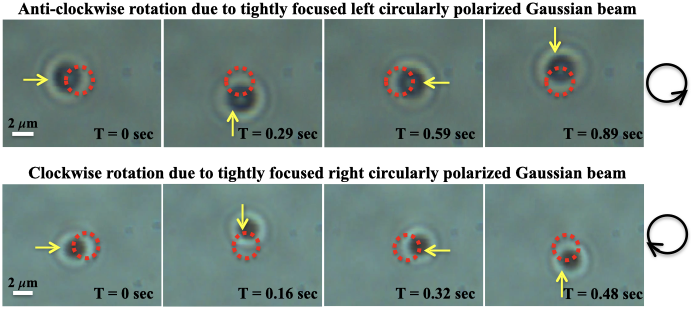}
    \caption{Time-lapsed frames of a video recording showing the rotation of particles by tightly focused circulation polarized light. The red circle marks the trajectory of the particle and yellow arrow indicates the position in that time frame. The rightmost circular panel show the orientation of particle's movement.}
    \label{fig:particlerotate}
\end{figure}

\begin{figure}[h]
    \centering    \includegraphics[width=0.75\textwidth]{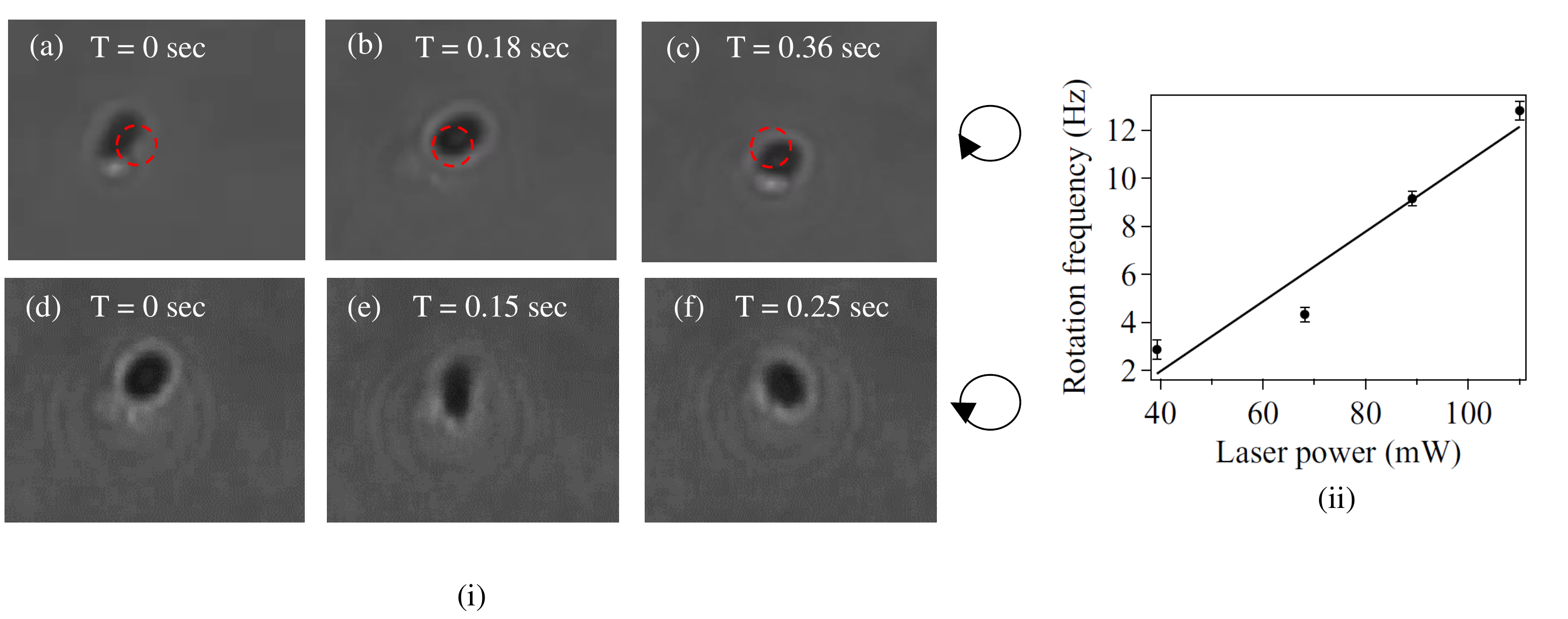}
    \caption{Time-lapsed frames of a video recording showing the rotation of particles when the input light helicity is changed during the experiment. Figs. (a) to (c) denote rotation around the beam axis with input RCP, while (d) to (e) denote rotation around the center of mass with input LCP. The red circle marks the trajectory of the particle in (a)-(c). The rightmost circular panel show the orientation of particle's movement for (a)-(c) and (d)-(f). (ii) Rotational frequency (around the beam axis) of a particle as a function of input laser power. }
    \label{fig:changingqwp}
\end{figure}

We now attempt to observe these effects experimentally on microscopic birefringent probe particles to verify our simulations. We use a conventional optical tweezers setup comprising of  an inverted microscope (Olympus BX71) with an oil-immersion 100X objective (Olympus, NA 1.3), and a high power diode laser (1064 nm, 500 mW) coupled into the back port of the microscope. We control the polarization of light using a quarter wave plate (QWP), and change the helicity of the beam by rotating the QWP by $90^o$.  For the probe particles, we use RM257 vaterite liquid crystal particle, which are optically anisotropic and  birefringent so as to transfer angular momentum from the beam into the particles, and thus probe the effects of TM and LSAM. We  construct the sample chamber using a glass slide and a cover-slip of RI 1.814, into which we add around $20 \mu$l of the aqueous dispersion of RM257 particles \cite{sandomirski2004highly}, which are elliptical, and of mean size $2~\times~1~\mu m$. We ensure that the incident power is constant for both left and right circular polarizations to avoid any difference in trapping conditions.\\
\newline
On coupling the laser into the microscope, we observe the formation of concentric off-axis intensity rings (see Supplementary Information, Fig. S1) around the beam center similar to that displayed in Fig.~\ref{fig:ElectricSpin}. The image of the rings is optimized by changing the $z$ $-$ focus of the microscope (imaging performed with transmitted light from the microscope lamp coupled into a camera attached to the side port of the microscope). As a result, particles tend to assemble in the ring as we reported in Ref.~\cite{basudevpra2013}. However, when we trap a single particle at the appropriate $z$-depth, and adjust the microscope focus to image it adequately, we are able to observe rotation of the particle around the beam axis (see Videos 1 and 2 in the Supplementary Information) - either clockwise or anticlockwise as displayed in  time-lapsed images in Fig.~\ref{fig:particlerotate}. The trajectories of the particle are shown as red dotted circles. The first row shows rotation in an anti-clockwise direction. When we modify the orientation of the QWP by $90^o$, the rotation direction of particle flips (becomes clockwise), as shown in the second row of time lapsed images in Fig.~\ref{fig:particlerotate}. 
However, these are for two different particles in separate experiments. We have changed the orientation of the QWP for the same particle as well, but often lose the particle from the trap in that process. This is because  - the input helicity - or the Jones vector in Eq.~1 is modified, so that the  output electric field intensity in the sample plane is also modified - as a result of which the trap often becomes unstable. However, in a few experiments, we do manage to keep the particle trapped by rotating the QWP very slowly. One such case we report here - which is shown in Video 3, and in time-lapsed images in Fig.~\ref{fig:changingqwp}(i). In this case, we observe rotation of the particle both around the axis of the beam in the clockwise direction (input RCP) demonstrated in Figs.~\ref{fig:changingqwp}(a)-(c), and around the particle body axis in the anti-clockwise direction for input LCP as demonstrated in Figs.~\ref{fig:changingqwp}(d)-(f). Note that, while the former is due to the enhanced TM in our system,  the latter is invariably the demonstration of the LSAM, which increases over TM when the axial trapping depth is modified. We have verified on this in our description of  Fig.~\ref{fig:TSMLSAM}, where - as we change the $z$-distance from the focus, the TM and LSAM behave differently. We believe that even when we change the QWP slowly, the intensity at the trapping plane still changes enough to change in the equilibrium position of the trapped particle in the $\mathbf{z}$ direction. This is what results in the trapped particle sampling higher LSAM than TM (note that the LSM at $z~=~1~ \mu m$ is more than the TM, whereas at $z~=~2~ \mu m$, these are similar). This is also apparent in Video 3 (snapshots in Fig.~\ref{fig:changingqwp} - compare the upper and lower rows), where the particle seems to undergo a change in shape as we change the QWP - which clearly indicates that the $z$-depth is changed. The fact that the effects of TM and LSAM are different as the trapping depth is modified is thus apparent in our experiment. We also observe that the frequency of rotation increases as we increase power as is shown in Fig.~\ref{fig:changingqwp}(ii), which is expected as the magnitudes of both electric and magnetic field will increase as we increase the input intensity. We demonstrate these events in Videos 4-7 in the SI.

The crucial issue is to decide whether the canonical (orbital) momentum $\mathbf{P_o}$ is responsible for the rotation of particles around the beam axis. This, however, cannot be the case since there is no dependence on input helicity in $\mathbf{P_o}$, so the direction of rotation should be independent of the helicity of input light. Thus, we can conclude that the rotation we observe is solely due to the effect of the Belinfante spin, which makes our experiment perhaps the first observation of this erstwhile elusive quantity for propagating light fields. We would also like to point out here that it is not feasible that this rotation occurs due to azimuthally asymmetric scattering from the particle, which may lead to the generation of orbital angular momentum \cite{mondal2015generation}, since the particle comes to rest as soon we remove the QWP. 
\section{Conclusions}
We observe that the TM or the transverse Poynting vector of a tightly focused Gaussian beam (optical tweezers) is explicitly helicity-dependent, while the TSAM remains helicity independent. This dependence demonstrates spin-momentum locking for tightly focused Gaussian beams. In addition, the TM is zero at the beam center and increases off-axis as we introduce a stratified medium in the path of the beam and increase its RI contrast. Thus, when we use a cover slip with RI 1.814 (mismatched condition) for the sample chamber of the optical tweezers, we observe in our simulations that the transverse extent of both the TM and field intensity increase with axial distance from the beam focus. As a result, there is sufficient overlap of high intensity - which can facilitate optical confinement; and high TM - which can lead to the observation of rotational effects around the beam axis. We validate this observation experimentally with birefringent particles where we demonstrate both clockwise and anti-clockwise rotation around the beam axis for input RCP and LCP with cover-slips of RI 1.814. Interestingly, when we change the helicity in the course of an experiment, we sometimes observe a transition from rotation around the beam axis (in a particular sense) to rotation about the particle axis (in the opposite sense). This implies that the particle samples predominantly TM in one case and LSAM in the other case. This observation is also expected from our simulations which demonstrate that the relative magnitudes of TM and LSAM are modified at different axial distances from the trapping beam focus, so that as the axial position of the particle is slightly modified in the process of changing helicity using a QWP, either the TM or the LSAM dominate. The rotation around the beam axis is the clear manifestation of Belinfante spin, which is solely responsible for the spin-dependence of the Poynting vector, the canonical component being spin-independent. We believe this to be a simple but robust way to observe the effects of Belinfante spin, which have often proved to be rather elusive to experimental observation. We are presently in the process of extending our studies to the tight focusing of higher-order Gaussian beams (Hermite-Gaussian) including those carrying intrinsic orbital angular momentum (Laguerre-Gaussian) to determine more interesting and intriguing effects of the interaction between spin and orbital angular momentum in the presence of a stratified medium.
\section{Acknowledgments}
The authors acknowledge the SERB, Department of Science and Technology, Govt of India $($project no. $EMR/2017/001456)$ and IISER Kolkata, Mohanpur, India, for research funds. They also acknowledge Mr. Sourav Islam of IISER Kolkata for help in the experimental measurements, and Dr. Georgios Vasilakis of Foundation for Research and Technology-Hellas, Greece for critical comments.

\end{document}